\documentclass[
 ,draft            
 ,numberedheadings 
 ]
  {aipproc}

\layoutstyle{6x9}

\date{October 16, 2005}

\begin{document}

\title{Spiral Spin Order and Transport Anisotropy in Underdoped Cuprates}



\author{Valeri N. Kotov}{ address={
 Institute of Theoretical Physics,
 Swiss Federal Institute of Technology (EPFL),\\
CH-1015 Lausanne, Switzerland }}

\author{Oleg P. Sushkov}{
  address={School of Physics, University of New South Wales, Sydney 2052, Australia}
}

\begin{abstract}
We discuss the spiral spin density wave model   and its application
 to explain properties of underdoped La$_{2-x}$Sr$_x$CuO$_4$.
 We argue that the spiral picture is theoretically well
 justified in the context of the extended $t-J$ model, and then 
 show that it can explain a number of observed features, such as
 the location and symmetry of the incommensurate peaks in elastic neutron scattering, 
as well as the in-plane resistivity anisotropy. A consistent description
 of the low doping region (below 10\% or so) emerges from the spiral formulation, 
 in which the holes show no tendency towards any 
type of charge order and the physics is purely spin driven.
\end{abstract}

\maketitle


\section{Introduction}

A popular scenario to explain the complex physics
of the high-temperature superconductors is based on 
the  idea that  the ground state of these materials 
exhibits some form of charge ordering tendency 
 (charge stripes, checkerboard order, etc.)  \cite{Kivelson,Tranquada,Zhang},
 which, in turn, can lead to incommensurate magnetism (spin stripes). 
From the outset we  state that we do not subscribe to
 this point of view. In order  to illustrate the direction of our efforts, we 
will  take as an example the La cuprate family,
 of which the most studied representative is 
 La$_{2-x}$Sr$_x$CuO$_4$ (LSCO) ($x$ is the hole doping). 
 This compound is commonly believed to  show ``dynamic'' charge order near
 the special doping value $x = 1/8$, while static order
(although quite weak) has been observed with additional Nd co-doping,
 or upon the substitution
La$\rightarrow$Ba. However the rest of the LSCO phase diagram at low
 doping $x<1/8$  does not exhibit any charge order \cite{Padilla} 
 making it rather hard to accept the universal concept that holes
in antiferromagnets fundamentally tend to segregate into charge stripes 
or similar structures.
 
 We  start from the guiding principle that  spin order 
  is the primary phenomenon while  charge order is an exception
 to the rule and could possibly occur under special circumstances only
(such as dopings corresponding to commensurate spin structures).
 This leads to a natural candidate for the
 ground state - the spiral spin state, as pointed out quite a while ago
 \cite{SS}. We  will seek to provide  theoretical
 support for the validity of the spiral picture  within
 the extended $t-J$ model, as well as  
 explain two sets of phenomena observed in LSCO:
\begin{itemize}
\item The presence of incommensurate magnetic order
 (with incommensurability=doping) 
both in the metallic (superconducting) and insulating phases
 at low doping, and in particular the mysterious change
 of the incommensurability direction by 45$^{\circ}$ at
 the metal-insulator  transition point $x\approx0.055$ \cite{Fujita}.

\item The in-plane  transport anisotropy, as  high as 50\%,
 observed in   the insulating spin glass
 phase ($0.02<x<0.055$) \cite{Ando}.
\end{itemize}

The near equality of   incommensurability and doping is
 usually considered as one of the successes of the charge stripe
 scenario, whereas the 45$^{\circ}$ rotation could follow
from considerations based on hole-pair checkerboard order \cite{Zhang}.
The transport anisotropy  of course also could be interpreted as
 a signature of (fluctuating) charge order, although no specific calculations
 have been performed \cite{Kivelson}. 

In the spirit of our alternative
 philosophy we will show that  a theory based on holes moving in an antiferromagnet and causing
  the formation of a spiral spin density wave  can explain the
 above phenomena, giving in  particular a
{\it quantitative} value for the magnitude of the transport
 anisotropy. The spiral theory is  Fermi liquid in nature \cite{remark}, without
 any charge ordering tendencies, and stands on firm theoretical ground.
It also provides a unified and consistent picture of the relationship
 between incommensurate magnetism and transport anisotropy.

\section{Spiral Order: Stability and Change of Symmetry}

The spiral order is generated by the hole motion in the N\'{e}el antiferromagnetic state,
in an attempt of the magnetic background to partially relieve the
 frustration caused by the hopping \cite{SS}. This leads to the
 non-collinear configurations shown in Fig.~1.
Parametrizing the magnetic order as:
 $|\mbox{i}\rangle = \!  e^{i \theta({\bf{r_i}}){{\bf{m}}}
\cdot{\bf{\sigma}}/2}| \! \uparrow\rangle,
 |\mbox{j}\rangle =  \! e^{i\theta({\bf{r_j}}){\bf{m}}
\cdot{\bf{\sigma}}/2}| \! \downarrow\rangle$,
$\mbox{ i $\in$  ``up''  sublattice}, \mbox{   j $\in$  ``down''  sublattice}$,
the angle of deviation from collinearity is
$\theta({\bf{r_i}}) = {\bf Q}.{\bf{r_i}}$, where ${\bf Q}$ is the spiral vector 
directed along the $(1,1)$ or $(1,0)$ lattice directions.
 These are both  co-planar configurations, and
 the unit vector ${\bf{m}}$ is perpendicular to the spin plane.
For low doping corresponding to small  deviations from N\'{e}el order,
 one finds that ${\bf Q}$ is proportional to doping: ${\bf Q} = \frac{Zt}{\rho_{s}}
\ x \ [(1,1) \mbox{or} (1,0)]$. Here $t$ is the hopping, ${\rho_{s}}$ is the spin stiffness,
 and $Z$ is the  quasiparticle residue at the points $(\pm \pi/2, \pm \pi/2)$, corresponding
 to the minima  of the hole dispersion at low doping.

\vspace{0.4cm}
\begin{figure}[h]
\includegraphics[height=.22\textheight]{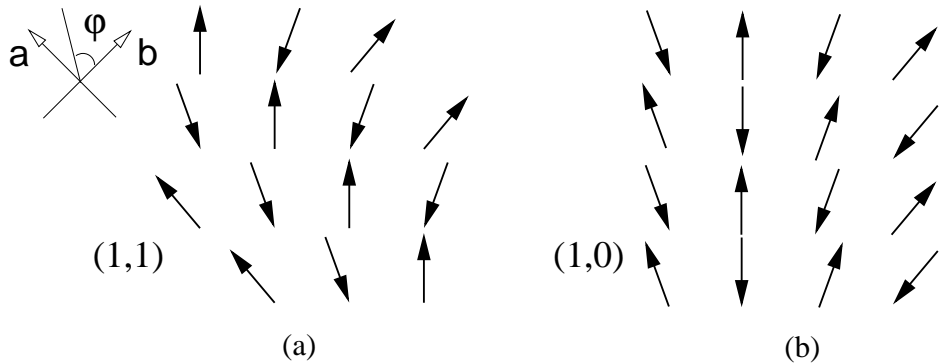}
\caption{Two types of spiral order on a square lattice.}
\end{figure}

It was realized shortly after the first several works on the spiral physics that  
 the spiral state  have a tendency towards some kind of instability
 in the charge sector (phase separation) \cite{Auerbach}. 
Indeed, within the $t-J$ model one finds  a negative charge
 compressibility, $\chi^{-1}<0$, defined, as in Fermi liquid theory,
through a derivative of the ground state energy $\chi^{-1}=\partial^{2}E/\partial x^{2}$.
 The presence of such an instability towards a hole segregated state
 would then mean that charge stripes in some form are likely
 to be present in the ground state. Under what conditions would
 the Fermi liquid physics survive?
 A possibility that we have recently explored in detail is the
 presence of additional (next nearest neighbor) hoppings  $t',t''$ \cite{Kotov1}.
 For LSCO the values of these parameters are quite small:
 $t'/J \approx -0.5, t''/J \approx 0.3$, where $J$ is the magnetic exchange,
 and $t/J \approx 3$ \footnote{We set $J=1$ from now on.}. 
 However we find that their presence is crucial
 for the stability of the system. 
 In order to perform as accurate calculations as possible
 we follow a two-step procedure: (1.) The one hole properties,
 such as $Z$ and the low-energy dispersion 
$\epsilon_{\bf k}\approx \frac{\beta_1}{2}k_1^2+\frac{\beta_2}{2}k_2^2$
near the nodal points are calculated in the self-consistent Born
 approximation, and (2.) These are inserted into the many-body fermion-magnon
low-energy  vertices which are then treated in perturbation theory
(loop expansion) in powers of  $Zt$.
We call this technique ``chiral  perturbation theory'', by analogy with QCD.
The perturbative parameter is not small, since  $Z \approx 0.34$, and thus
$Zt \sim 1$. However we find that the perturbation theory converges
numerically extremely well \cite{Kotov1}, as if governed by the effective
 coupling constant $g_{eff} \approx (Zt)^{2}/\pi \approx 0.3$.  
\begin{figure}[t]
\includegraphics[height=.31\textheight]{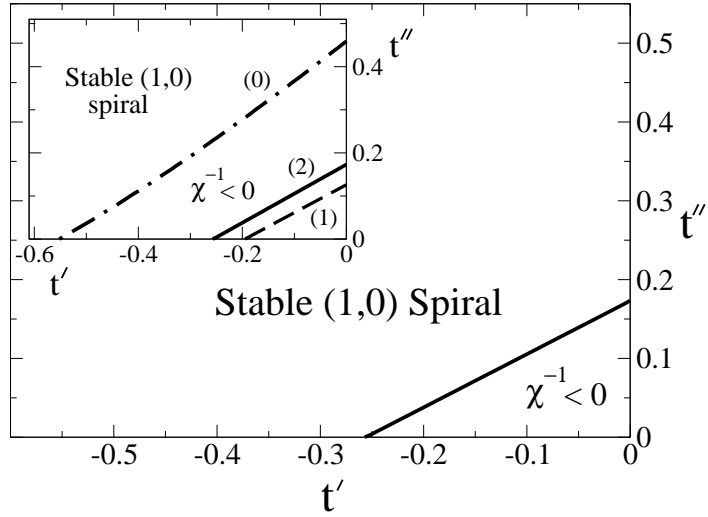} 
\caption{Stability diagram in the metallic phase at low doping.
$t',t''$ are in units of $J$. Inset: Stability boundary at different
 levels of approximation (loops), showing the good convergence.}
\end{figure}
The results, presented in Fig.~2, indicate that   fairly small values  
of  $t',t''$ stabilize the uniform $(1,0)$ spiral state. The inset shows
 the transition  line calculated in different loops (powers of $(Zt)^{2}$),
confirming the good convergence and reliability of the results.
Our results are in a way not surprising because  DMRG work \cite{White}
 had  found that charge stripes which are stable at $t'=t''=0$ become
 unstable (and thus the system becomes uniform) upon introduction
 of $t'$. In our theory the stability  depends physically on the
 shape of the Fermi surface -- while at  $t'=t''=0$
it is very elongated in the antinodal direction ($\beta_2 \ll \beta_1$),
 upon introduction of  $t',t'' \neq 0$ it becomes more spherically
 symmetric (for example $\beta_1 \approx \beta_2 \approx 2.2 $ for
 the LSCO values), and this effective two-dimensionality leads to
 the stability of the uniform spiral state. It should be noted that
 our calculations are valid for small doping $x \ll 1$, since
 we keep track of only  the terms of order $x^{2}$ in the ground state
 energy, and thus the compressibility has the expansion
$\chi^{-1} = c_{0} +  c_{2} (Zt)^{2} + c_{4} (Zt)^{4} + \dots$,
 leading to a stability line which does not depend on doping.

 The ground state energy change (relative to the undoped N\'{e}el state)
 due to the spiral formation with the two  symmetries, 
 satisfies the relation:
$\Delta E_{(1,1)} = 2 \Delta E_{(1,0)} \equiv \chi^{-1} x^{2}$. 
 This formula is correct to all perturbative orders that we have checked
 and can be traced to the different number of occupied hole
 pockets for the different spiral orientations \cite{Kotov1}.
 Above the stability boundary where $\chi^{-1}>0$, the $(1,0)$ state has lower
 energy\footnote{It may seem surprising that $\Delta E_{(1,0)}>0$, but one can show
 that the overall energy change relative to the {\it doped} collinear (N\'{e}el) 
state is negative \cite{Kotov1}.}. The $(1,0)$ symmetry is the correct one
 for the metallic phase of LSCO ($x>0.055$), where the elastic
 neutron scattering peaks are in a ``parallel'' pattern around
 $(\pi,\pi)$ \cite{Fujita}.
 What happens if the system becomes an insulator?
 As is clear from the calculation of the energy \cite{Kotov1}, the
 $(1,0)$ state has lower energy only due to the Fermi motion energy
$E_{F}$. On the other hand if we let $E_{F} \rightarrow 0$, which
 would be the case at the transition to  the spin glass region $x<0.055$ where the holes
 are localized,  then the $(1,1)$ state is selected
 (consistent with neutron peaks in ``diagonal'' pattern \cite{Fujita}). Of course this
 argument can be applied only up to the metal-insulator transition point,
 beyond which a detailed model for the spiral formation has to be constructed
\cite{Sushkov}.
 Nevertheless it is clear that the spiral model can explain correctly
 the presence  and the symmetries of the elastic neutron scattering peaks
 both in the metal and in the insulator. The exact location
 of those peaks (determined by ${\bf Q}$) is also in very good
 agreement with experiment \cite{Sushkov}.

\section{Transport Anisotropy induced by Spiral Order}
 We now turn to  the insulating spin-glass region
 $0.02<x<0.055$, where the presence of incommensurate magnetic peaks
(with $(1,1)$ symmetry)  is clearly related to the in-plane DC 
resistivity anisotropy $\rho_{b}/\rho_{a}$ \cite{Ando} ($\hat{a}$ and $\hat{b}$
are the orthorhombic coordinates).
From elastic neutron scattering \cite{Fujita} the incommensurability
is determined to be along the orthorhombic $\hat{b}$ direction, meaning
 that in the spiral picture it is in the $(1,1)$ direction, as shown in
 Fig.~1(a). Experimentally $\rho_{b}/\rho_{a} \approx 1.5$
at the lowest temperature $T \approx 10 {\mbox K}$ and then decreases,
 disappearing completely  around $100  {\mbox K}$ where the system
 becomes quasi-metallic.
\begin{figure}[t]
\includegraphics[height=.31\textheight]{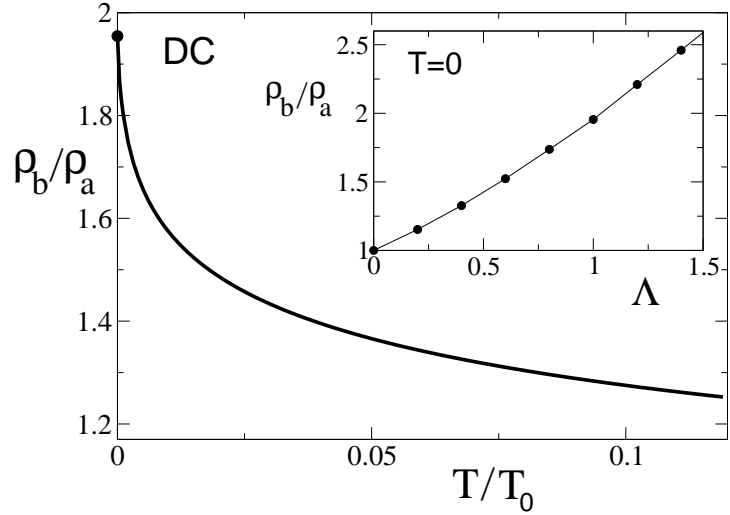}
\caption{In-plane DC resistivity anisotropy in the variable-range
 hopping regime ($\Lambda=1$).
 Inset: Maximum ($T=0$) anisotropy as a function of $\Lambda = \frac{Z^2t^2}{\pi \beta \rho_s}$.}
\end{figure}
                                                                                                         
Since it is clear that the largest anisotropy is accumulated in
 the low-temperature region, we will concentrate on the
strongly-localized, variable-range hopping (VRH) temperature range
  below approximately $30 {\mbox K}$. The resistivity of LSCO
 is well fit by the Mott 2D VRH formula $\rho \sim \exp{(T_0/T)}^{1/3}$,
 with characteristic values $T_0 \approx 200 {\mbox K} - 500 {\mbox K}$,
depending on doping. In order to address the anisotropy problem theoretically,
 one needs to develop a theory of spiral formation arising from the
 presence of (randomly distributed) localized holes. Recent works  have developed
such theories  \cite{Sushkov,HCS}, and in particular
 we have shown \cite{Sushkov}  that holes, localized around  $Sr$ ions can 
``carry'' spiral correlations with them that decay in dipolar fashion at
long distances. At finite doping the dipoles can order, 
producing a spiral state as the one in   Fig.~1(a).
Within this framework we have calculated the VRH conductivity anisotropy
by calculating first the wave-function of a localized hole
(with localization length $1/\kappa$), in the presence of spiral correlations \cite{Kotov2}:
$\psi(\kappa r,\varphi)=\psi_{0}(\kappa r) + \psi_{2}(\kappa r) \cos(2\varphi)  +
\psi_{4}(\kappa r) \cos(4\varphi) + \dots$. This wave-function is
 anisotropic (the angle $\varphi$ is defined in Fig.~1(a)), which is physically
 natural since the hole ``feels'' a different environment depending on the
 lattice direction, due to the non-collinearity of the spiral state.
The strength of the anisotropy is controlled by the
 parameter  $\Lambda = \frac{Z^2t^2}{\pi \beta \rho_s}$, and
$\Lambda \approx 1$  for LSCO (where $\beta \equiv \beta_{1} \approx \beta_{2}\approx 2.2$).  
The overlap of the wave-functions discussed above leads to the resistivity
 anisotropy, shown in  Fig.~3 \cite{Kotov2}. The magnitude of the anisotropy
 agrees extremely well with experiment \cite{Ando}. We emphasize that
 there are no adjustable parameters in our theory, although it is certainly
 valid at low doping only (and, similarly to the stability results of Section 2, the
 curve in Fig.~3 is doping independent). Perhaps most importantly,
 our analysis shows that the transport anisotropy can be due to the
underlying  incommensurate magnetic correlations, rather than a tendency
 of the charges to self-organize.                                                                                                                 

\section{Discussion and Outlook}

 The spiral spin density wave theory passes both the fundamental
 and phenomenological tests: it is well theoretically supported 
in the context if the extended $t-J$ model, and is capable of
 describing  magnetic and transport properties of LSCO.
 A novel finding is that the in-plane transport
 anisotropy fits well into the spiral picture on a quantitative level --
 we  consider this result particularly important because 
the anisotropy is a rather elusive quantity and usually hard to calculate
 consistently. 
 We should mention that the reported work essentially  explored 
only the structure of the ground state; how consistent the
 spiral picture would be with experiment at higher energy
 is not yet clear. Nevertheless all low-temperature  
 LSCO experiments we have looked at so far lead us to the conclusion
 that the physics at low doping is {\it spin driven}.  
The recently observed ``magic'' doping concentrations, where
the conductivity of LSCO shows dips \cite{Magic}, are also due,
 in our view, to spin related phenomena (such as special points
 where the spin structure becomes commensurate). 
Currently research is in progress to  explore further the predictions of the
 spin density wave approach (e.g. in the direction of including
 lattice effects and studies of magnetotransport). The theory 
provides, we believe, a consistent picture of the complex interplay
between spin and charge dynamics in the underdoped region of the cuprates.

\vspace{0.3cm}
V.N.K. acknowledges the financial support of the Swiss National Fund.

\bibliographystyle{aipproc}   


\begin{thebibliography}{9}
\bibitem{remark}
To avoid misunderstanding we stress that this is a Fermi liquid
 with a small Fermi surface.
\end{thebibliography}




\end{document}